\documentclass[aps,prl,twocolumn,superscriptaddress]{revtex4}
\usepackage{amsmath,amssymb,graphicx,color,braket,hyphenat,makeidx,hyperref}

\newcommand{\ketbras}[2]{\ket{#1}\!_{#2}\!\bra{#1}}
\newcommand{\ketbrast}[3]{\ket{#1}\!_{#2}\!\bra{#3}}
\newcommand{\cg}{r_c}
\newcommand{\rg}{r_r}
\newcommand{\hg}{r_h}
\newcommand{\ce}{\bar{r}_c}
\newcommand{\re}{\bar{r}_r}
\newcommand{\he}{\bar{r}_h}
\newcommand{\cG}{\Gamma_c}
\newcommand{\rG}{\Gamma_r}
\newcommand{\hG}{\Gamma_h}

\begin{document}

\title{A small quantum absorption refrigerator with reversed couplings}

\author{Ralph Silva}\affiliation{H. H. Wills Physics Laboratory, University of Bristol, Tyndall Avenue, Bristol, BS8 1TL, United Kingdom}
\author{Paul Skrzypczyk}\affiliation{ICFO-Institut de Ciencies Fotoniques, Mediterranean Technology Park, 08860 Castelldefels, Barcelona, Spain}
\author{Nicolas Brunner}\affiliation{D\'epartement de Physique Th\'eorique, Universit\'e de Gen\`eve, 1211 Gen\`eve, Switzerland}

\begin{abstract}
Small quantum absorption refrigerators have recently attracted renewed attention. Here we present a missing design of a two-qubit fridge, the main feature of which is that one of the two machine qubits is itself maintained at a temperature colder than the cold bath. This is achieved by 'reversing' the couplings to the baths compared to previous designs, where only a transition is maintained cold. We characterize the working regime and the efficiency of the fridge. We demonstrate the soundness of the model by deriving and solving a master equation. Finally, we discuss the performance of the fridge, in particular the heat current extracted from the cold bath. We show that our model performs comparably to the standard three-level quantum fridge, and thus appears appealing for possible implementations of nano thermal machines.
\end{abstract}

\maketitle

\twocolumngrid

\section{Introduction}

A considerable amount of work has been devoted to the study of quantum thermal machines. First works date back to the study of the thermodynamics of lasers \cite{schulz,geusic}. More recent works discussed quantum thermal machines as a platform to explore quantum thermodynamics \cite{book}, while others aimed at proposing experimentally feasible designs for thermal machines at the nano scale, see e.g. \cite{sanchez,bjorn,esposito,abah,nernst,altintas,dechant}. 

A class of quantum thermal machines that received recently great interest are autonomous, or self-contained, thermal machines; see \cite{review1,review2} for recent reviews. The specificity of these machines is that they function without any external source of work, but simply via thermal contact to heat baths at different temperatures. In particular, small quantum absorption refrigerators use only two thermal reservoirs, one as a heat source, and the other as a heat sink, in order to cool a system to a temperature lower than that of either of the thermal reservoirs. 

Moreover, the appeal of these machines resides in their simplicity. Few quantum levels are enough to construct a quantum absorption refrigerator. The simplest designs, i.e. the smallest possible machines consist of a single qutrit (i.e. a 3-level system) \cite{schulz,linden10}, 2 qubits \cite{palao,levy} or 3 qubits \cite{linden10,skrzypczyk11,brunner12,correa13}.  Other designs were discussed as well \cite{GevKos96,LevKos12,luis,gelbwaser,bellomo}. The efficiency of such machines was investigated, and certain designs were proven to achieve Carnot efficiency \cite{schulz,skrzypczyk11}. Moreover, the significance of quantum effects in some of these machines, in particular entanglement \cite{brunner14} and coherence \cite{uzdin,marcus}, was discussed. The effect of applying squeezing to the heat baths was studied in \cite{rossnagel,correa2}. Finally, schemes for experimental implementations of these ideas were proposed \cite{mari,venturelli,chen}.

Interestingly, the functioning of essentially all these quantum refrigerators can be captured by the following idea. The cooling process can be divided in two steps. First, a transition inside the machine (usually resonant with the system to be cooled) is engineered in such a way that its temperature, a so-called virtual temperature \cite{brunner12}, is lower than the temperature of the coldest bath. Second, this transition is then (thermally) coupled to the system, which thus thermalizes towards the virtual temperature (or at least to a temperature colder than the coldest bath). 

In this work, we present a novel design for a quantum absorption refrigerator consisting of only two qubits. The specificity of our model is that cooling is achieved via `reversed couplings' compared to previous designs. More precisely, whereas previous designs function by preparing a virtual temperature, our design allows for the direct cooling of a physical qubit. Loosely speaking, instead of cooling a single transition (i.e. a virtual qubit), we cool a physical qubit. Furthermore, the way in which the fridge is coupled to the thermal reservoirs is also in contrast with previous designs. Instead of connecting each qubit to a thermal thermal reservoir (see e.g. \cite{linden10}), we now connect the reservoirs to transitions within the joint state of the two qubits. We characterize the working regime and the efficiency of the fridge, and demonstrate the soundness of the model by deriving and solving a Master equation which models the dynamics in a weak coupling regime. Finally, we discuss the performance of our model, using as a figure of merit the heat current extracted from the cold bath. Interestingly, we find that our model is comparable to the standard three-level quantum absorption refrigerator \cite{schulz,linden10,correa2}, and extracts a greater amount of heat under some conditions. We believe that this makes our model appealing for future implementations of nano thermal machines, as well as suggesting the use of reverse couplings as a viable alternative to the usual construction of these machines.

\section{The Model}

The fridge is comprised of two qubits. The first qubit, with levels $\ket{0}_1$ and $\ket{1}_1$, and energy gap $E_1$ represents the object to be cooled. The second qubit has levels $\ket{0}_2$ and $\ket{1}_2$, and energy gap $E_2$. The free Hamiltonian of the system is thus given by 
\begin{equation}\label{H0}
	H_0 = E_1 \ketbras{1}{1} \otimes \hat{\mathbb{I}}_2 + E_2 \; \hat{\mathbb{I}}_1 \otimes \ketbras{1}{2}
\end{equation}
where $\hat{\mathbb{I}}_j$ denotes the identity operator for qubit $j$. By design of the machine, we choose $E_2>E_1$.

Next, we introduce two thermal reservoirs: the heat source reservoir at temperature $T_h$ and the sink (or room) reservoir at temperature $T_r<T_h$. In the following, we will mainly use the inverse temperatures $\beta_h = 1/k_B T_h$ and $\beta_r = 1/k_B T_r$. 

The system of two qubits (that is referred to as the fridge henceforth) is coupled to the thermal reservoirs in the following way. The sink bath is selectively coupled to the energy gap $E_2+E_1$. Thus the fridge can transition between the states $\ket{00}=\ket{0}_1 \ket{0}_2$ and $\ket{11}=\ket{1}_1 \ket{1}_2$ by absorbing or emitting an energy quanta of $E_2+E_1$ from or to the sink reservoir. Second, the hot bath is selectively coupled to the energy gap $E_2-E_1$, allowing the fridge to transition between the states $\ket{10}=\ket{1}_1 \ket{0}_2$ and $\ket{01}=\ket{1}_1 \ket{0}_2$ by absorbing or emitting an energy quanta of $E_2-E_1$. Additionally, the first qubit of the fridge is coupled to a third reservoir at temperature $T_c \leq T_r$. Hence the first qubit can absorb or emit energy quanta of $E_1$ with this `cold' reservoir.

\begin{figure}[b]
\includegraphics[width=\linewidth]{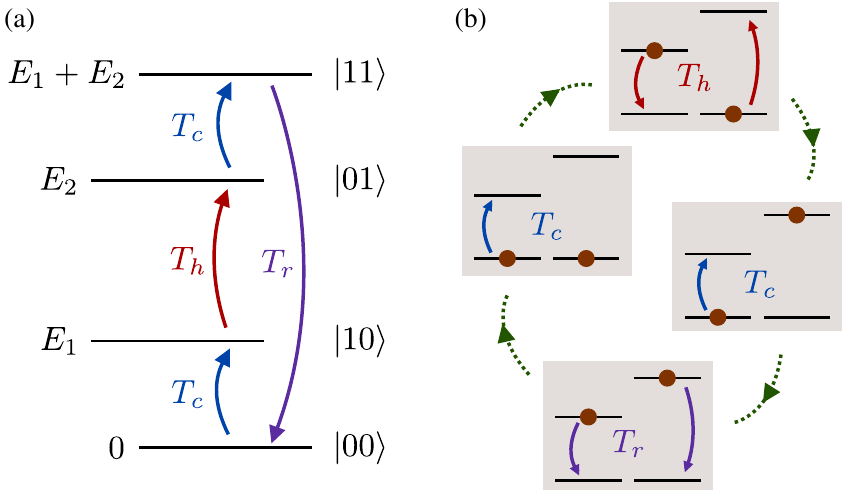}
\vspace{-10pt}
\caption{\label{4level} (a) The 2 qubit fridge viewed as a 4-level system, with the couplings to the hot, cold, and sink reservoirs. (b) The four strokes of the cooling cycle. In each sub figure the balls represent the state each qubit can initially be found in, and the arrows depict the transitions that occur, driven by the respective baths. By making sequentially the four transitions shown (i.e. the four strokes), the system undergoes a cyclic evolution and completes one cooling cycle.}
\vspace{-20pt}
\end{figure}

The working model is summarized in Fig.\ref{4level} (a) which depicts the four levels of the system and the various transitions that are coupled to the reservoirs. In order to demonstrate cooling, the fridge must extract heat from the coldest reservoir, at $T_c$. Equivalently, if the fridge possesses a steady state of operation, the (reduced) state of the first qubit should be found at a temperature that is lower than $T_c$, which would predispose the cold reservoir to transfer heat \emph{to} rather than from the fridge. 

\section{Cooling cycle}

We start our discussion of the above model by describing the ``cooling cycle" of the fridge. From Fig. \ref{4level} (a), we observe that the thermal couplings have been chosen in order to allow the fridge to move through the following unique cycle, involving every level of the fridge:
\begin{equation}
	\ket{00}\rightarrow\ket{10}\rightarrow\ket{01}\rightarrow\ket{11}\rightarrow\ket{00}
\end{equation}

It is insightful to understand the cooling cycle as a four-stroke process, as illustrated in Fig. \ref{4level} (b). If one considers the fridge to begin in the state $\ket{00}$, the energy exchanged with the thermal reservoir in each stroke is as follows:
\begin{itemize}
	\item $\ket{00}\!\rightarrow\!\ket{10}$. Qubit 1 absorbs $E_1$ from the cold reservoir.
	\item $\ket{10}\!\rightarrow\!\ket{01}$. Qubits 1 and 2 together absorb $E_2-E_1$ from the hot reservoir.
	\item $\ket{01}\!\rightarrow\!\ket{11}$. Qubit 1 once again absorbs $E_1$ from the cold reservoir.
	\item $\ket{11}\!\rightarrow\ket{00}$. Qubits 1 and 2 together dump $E_2+E_1$ into the sink reservoir.
\end{itemize}

The net effect of the cooling cycle is that the fridge absorbs energy $2E_1$ from the cold reservoir and energy $E_2-E_1$ from the hot reservoir, and dumps the total energy $E_2+E_1$ into the sink reservoir.

Indeed, one should also consider the reversed cycle. Here the fridge \emph{emits} energy $2E_1$ to the cold reservoir. Now in order to determine whether cooling occurs, i.e. whether the fridge extracts heat from the cold reservoir, one must determine which of the two cycles, the cooling cycle or its reverse, is more likely to occur. To find out, we compute the entropy change associated to each cycle. Following the second law of thermodynamics, the preferred cycle is the one that tends to increase the total entropy of the system (fridge + reservoirs).

The entropy change of each reservoir is given by $\Delta S_i = \beta_i Q_i$, where $Q_i$ is the heat absorbed by the $i^{th}$ reservoir. Summing the contribution from each of the four strokes of the cooling cycle, one obtains the total entropy change
\begin{equation}\label{entropy}
	\Delta S = - \beta_c E_1 - \beta_h (E_2-E_1) - \beta_c E_1 + \beta_r (E_2+E_1)
\end{equation}
The entropy change of the reverse cycle is of course $-\Delta S$. Thus the condition that the fridge achieves cooling is simply that $\Delta S>0$, or equivalently 
\begin{equation}\label{virtualtemp}
	\beta_V = \frac{\beta_r (E_2+E_1) - \beta_h (E_2-E_1)}{2 E_1} > \beta_c
\end{equation}
where $\beta_V = 1/k_B T_V$, and $T_V $ is labelled the virtual temperature of the fridge qubit \cite{brunner12}. We shall show below that much of the working of the fridge may be understood from this virtual temperature. Notably, we see already that one can re-express the condition that the fridge achieves cooling by the condition $T_V<T_c$. That is, the virtual temperature of the fridge should be lower than the temperature of the cold bath for cooling to occur.

\section{Efficiency and Carnot point}

Our next focus is the efficiency of the fridge, defined as the ratio between the amount of heat drawn from the cold reservoir and the amount of heat drawn from the hot reservoir. More formally, the efficiency is given by
\begin{equation}
	\eta = \frac{Q_c}{Q_h}
\end{equation}
where $Q_h$ and $Q_c$ denote the heat currents \emph{from} the hot and cold reservoirs to the system. Similarly, we define the heat current \emph{into} the sink reservoir to the system as $Q_r$. Since the fridge implements a cycle, we can determine the ratios of heats exchanged from the relevant amounts exchanged in a single cycle: 
\begin{align}
	Q_h:Q_c:Q_r \; :: \; E_2-E_1:2E_1:E_2 + E_1 
\end{align}
which therefore determines the efficiency
\begin{equation}
\eta = \frac{2E_1}{E_2-E_1}
\end{equation}
Alternatively, one may use the virtual temperature in order to express the efficiency as
\begin{equation}\label{efficiency}
	\eta = \frac{ \beta_r - \beta_h }{ \beta_V - \beta_r } < \frac{ \beta_r - \beta_h }{ \beta_c - \beta_r },
\end{equation}
where the inequality follows from the cooling criterion $T_V<T_c$. Hence we obtain an upper bound for the efficiency given by  
\begin{equation}
\eta_C = \frac{ \beta_r - \beta_h }{ \beta_c - \beta_r }
\end{equation}
which is simply the Carnot efficiency for an absorption refrigerator working with the temperatures $T_c$, $T_r$, and $T_h$. Thus the Carnot point is approached when the virtual temperature approaches to the temperature of the cold reservoir, i.e. $T_V\to T_c$. From \eqref{entropy} - \eqref{virtualtemp}, one finds that in the limit, the entropy change $\Delta S \to 0$ and thus the cooling cycle approaches reversibility, as it must do. We will see later, when we look at a more detailed model, that  the heat currents also vanish in this limit, showing that we enter the quasi-static regime as $T_V\to T_c$, again as is necessary for reversibility.

\section{Detailed model of the fridge}

We now present a detailed model of the fridge. We will derive the steady state of the fridge, and discuss its cooling regime, the Carnot point, and operating points away from Carnot.

As we discussed above, several transitions within the two qubit system are coupled to thermal baths at different temperatures. Here each bath is represented by an (infinite) collection of qubits, with energy corresponding to the associated transition in the fridge. Thus the state of a qubit from bath $j \in \{ c,r,h \}$ is given by

\begin{equation}
	\tau_j = e^{-\beta_j H_j} = r_j \ketbras{0}{r} + \bar{r}_j \ketbras{1}{j},
\end{equation}
where $r_j = 1-\bar{r}_j = (1+e^{-\beta_r E_j})^{-1}$. We have that $E_c = E_1$, $E_r = E_1+E_2$ and $E_h = E_2 -E_1$.

The interaction between the qubit bath and the fridge are modeled as follows. In a time interval $\delta t$, there is a small probability $p_j \delta t$ (where $j \in \{ c,r,h \}$), that a qubit from reservoir $j$ interacts with the fridge. The interaction is energy conserving, and described by the Hamiltonian that implements a swap between the thermal qubit and the corresponding transition within the fridge. For the sink and hot reservoirs we have
\begin{align}
	H_r &= g_r \left( \ketbrast{00}{s}{11} \otimes \ketbrast{1}{r}{0} + c.c. \right), \\
	H_h &= g_h \left( \ketbrast{10}{s}{01} \otimes \ketbrast{1}{h}{0} + c.c. \right).
\end{align}
The coupling to the cold bath is given by
\begin{equation}
	H_c = g_c \left( \ketbrast{0}{1}{1} \otimes \mathbb{I}_2 \otimes \ketbrast{1}{c}{0} + c.c. \right).
\end{equation}
Note that the cold bath couples directly to qubit 1.

To prevent anomalous heat flows between the reservoirs while they are both coupled to the fridge, we take the interactions to be very strong (i.e. all of the strengths $g_i$ are very high), but to act for a short time, so that the probability that the fridge is coupled to more than a single thermal qubit is negligible. Moreover, we assume that there are no memory effects from the reservoir, thus the effect of the interaction on the fridge is obtained by tracing out the thermal qubit at the end of the interaction. Assuming that the duration of the interaction between the fridge and a thermal qubit is uncertain over the time it takes to for a complete swap, the effect is calculated via the time-averaged map. For the case of the interaction with heat bath $j\in \{c,h,r\}$,
\begin{align}\label{dissipator}
	 \Omega_j(\rho)  
	 = \text{Tr}_j \left[ \frac{g_j}{2\pi}\int_{0}^{2\pi/g_j} e^{-iH_j t} \left( \rho \otimes \tau_j \right) e^{+iH_j t} dt \right]
\end{align}
where $\text{Tr}_j$ denotes the partial trace over the bath qubit $j$.
The maps that describe the effect of an interaction with a hot or cold thermal qubit are calculated analogously.

The above model leads to a simple master equation that determines the evolution of the fridge:
\begin{align}\label{master}
	\frac{d\rho}{dt} &= i[\rho,H_0] + \sum_{j\in\{c,h,r\}} p_j \left( \Omega_j(\rho) - \rho \right)
\end{align}
where the free Hamiltonian of the two-qubit fridge $H_0$ is given in equation \eqref{H0}.

Solving the above master equation, we find that the steady state of the fridge is a diagonal state \footnote{Note that one obtains the same steady state using the reset model of thermalization used in Ref. \cite{linden10}.}, with its diagonal elements given below:
\begin{align}\label{e:steady state}
	\rho_S &=  \frac{1}{D} \mathrm{diag} \begin{pmatrix}
				\cg \left( \frac{\cg\hg}{p_r} + \frac{\ce\rg}{p_h} \right) + \rg \left( \frac{\cg\hg}{p_c} + \frac{\ce\he}{p_c} \right) \\
				\ce \left( \frac{\cg\hg}{p_r} + \frac{\ce\rg}{p_h} \right) + \hg \left( \frac{\cg\re}{p_c} + \frac{\ce\rg}{p_c} \right) \\
				\cg \left( \frac{\ce\he}{p_r} + \frac{\cg\re}{p_h} \right) + \he \left( \frac{\cg\re}{p_c} + \frac{\ce\rg}{p_c} \right) \\
				\ce \left( \frac{\ce\he}{p_r} + \frac{\cg\re}{p_h} \right) + \re \left( \frac{\cg\hg}{p_c} + \frac{\ce\he}{p_c} \right)
			\end{pmatrix}, 
\end{align}	
where		

\begin{equation}
	D = \left( \frac{1}{p_c} + \frac{1}{p_r} \right) \left( \cg\hg + \ce\he \right) + \left( \frac{1}{p_c} + \frac{1}{p_h} \right) \left( \cg\re + \ce\rg \right).
\end{equation}

Of course, one may take a different approach to the process of thermalization than the one we follow, which would lead in general to different dynamics. Interestingly, if one considers the baths to be collections of bosons instead, one arrives at a steady state that is identical to ours under the appropriate choice of thermal couplings (see Appendix). Also, if one uses the simpler `reset' model\cite{linden10}, one arrives at the same steady state.

We first discuss the condition for cooling. Since the cold qubit of the fridge is directly coupled to the cold reservoir, cooling occurs when the cold qubit of the fridge is maintained at a lower temperature than that of the reservoir; i.e.  its ground state population $r_1$ is larger compared to a qubit from the cold reservoir.
We find that
\begin{equation}\label{coolingcondition}
	r_1 = \cg + \frac{2(\ce^2\rg\he - \cg^2\re\hg)}{p_c D},
\end{equation}
Since $D>0$, the cooling condition $r_1>r_c$ reduces to
\begin{equation}
	\ce^2\rg\he - \cg^2\re\hg > 0 \;\;\;\;\;\Longrightarrow\;\;\;\;\; T_V<T_c,
\end{equation}
with the virtual temperature $T_V$ as defined in \eqref{virtualtemp}. Hence we recover the cooling condition derived above from simple entropic considerations. This demonstrates that the cooling condition is independent of the coupling parameters, but determined solely by the virtual temperature $T_V$. The latter depends only upon the static design of the fridge, i.e. the energies of the fridge qubits and the temperatures of the thermal reservoirs.

Another point of interest is the steady state of the fridge when operated at the Carnot point, i.e. $T_V=T_c$. In this case, the steady state factorizes into the tensor product state
\begin{equation}
	\rho = \begin{pmatrix} \cg & 0 \\ 0 & \ce \end{pmatrix} \otimes \begin{pmatrix} \frac{\cg\hg}{\cg\hg+\ce\he} & 0 \\ 0 & \frac{\ce\he}{\cg\hg+\ce\he} \end{pmatrix}
\end{equation}
Thus the cold qubit is in equilibrium with the cold reservoir, while the second qubit is at an effective temperature that lies between the temperatures of the hottest and coldest reservoirs. It is easily verifiable that the effective temperatures of the transitions between the pairs of energy levels $\ket{00}\leftrightarrow\ket{11}$ and $\ket{10}\leftrightarrow\ket{01}$ are equal to $T_r$ and $T_h$ respectively, and thus at the Carnot point, the fridge is in thermal equilibrium with every reservoir, in the sense that every transition that is coupled to a bath is stationary with virtual temperature equal to the bath temperature it is in contact with.

\begin{figure}[b!]
\includegraphics[width=\linewidth]{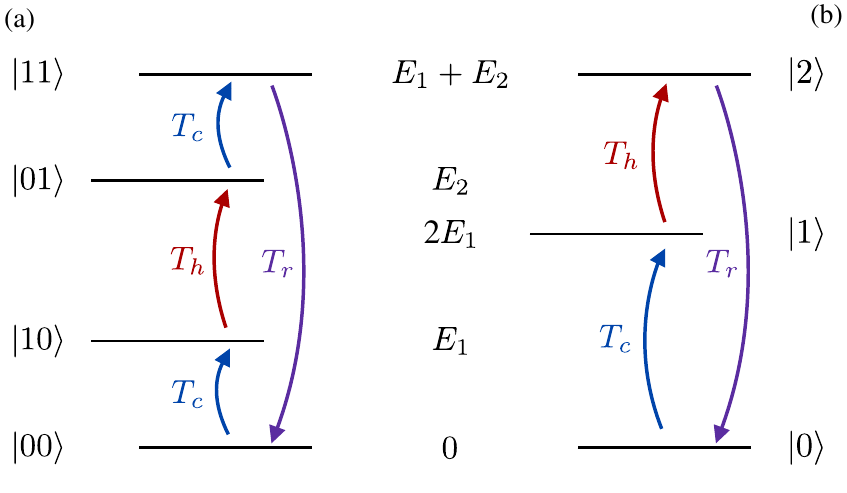}
\vspace{-10pt}
\caption{\label{sketch} Construction of equivalent qutrit fridge to compare performance.}
\end{figure}

\section{Performance of the fridge}

Here we discuss the performance of our fridge. As a figure of merit we consider the heat current extracted from the cold bath, calculated via the dissipator from \eqref{dissipator},
\begin{align}
	Q_c	&= p_c E_1 \text{Tr} \left[ \left( \Omega_c(\rho) - \rho \right) \ketbras{1}{s} \right] \\
	&= \frac{2(\ce^2\rg\he - \cg^2\re\hg)}{D}
\end{align}

\begin{figure*}[t!]
\includegraphics[width=1\linewidth]{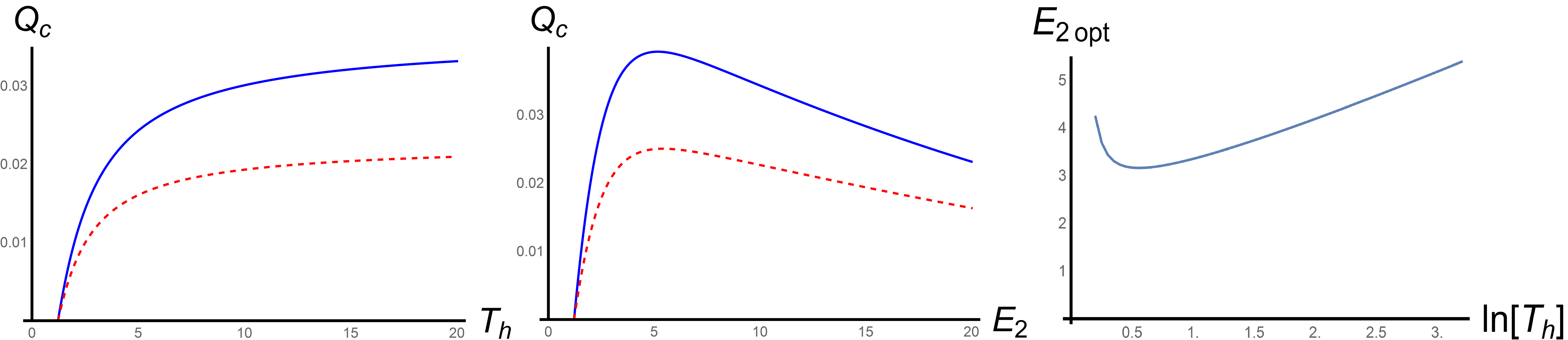}
\caption{\label{triple}Performance of the fridge (solid blue lines), and comparison with the three-level fridge (dashed red lines); see main text. (a) Heat current $Q_c$ drawn from the cold bath as a function of the temperature of the hot bath $T_h$. (b) $Q_c$ as function of the energy of the second qubit $E_2$, for fixed $T_h=20$. Clearly, the heat current is maximized at a finite value $E_2^{opt}$. (c) Dependency of $E_2^{opt}$ as a function of $T_h$. Finally, we note that in both (a) and (b), our model outperforms the standard three-level fridge.}
\end{figure*}

We proceed with a few examples. Without loss of generality, we set $E_1=1$ (hence setting the energy scale). As a case study, we choose $T_c=1$, $T_r=1.1$, and fix the coupling parameters: $p_c=p_r=p_h=1$. We first investigate the dependency of $Q_c$ with respect to the temperature of the hot bath $T_h$. The results are plotted (solid blue line) in Fig. \ref{triple}(a). The heat current increases monotonically with $T_h$, asymptotically approaching a saturation value
\begin{equation}
	\lim_{T_h\rightarrow\infty} Q_c = \frac{\ce^2\rg - \cg^2\re}{\frac{1}{2} \left( \frac{1}{p_c} + \frac{1}{p_r} \right) + \left(\frac{1}{p_c}+\frac{1}{p_h}\right)(\cg\re+\ce\rg)}.
\end{equation}

Second, we investigate the dependence of $Q_c$ with respect to $E_2$, fixing $T_h=20$. The results are plotted (solid blue line) in Fig. \ref{triple} (b). Interestingly, we observe that there is an optimal value for $E_2$, denoted $E_2^{opt}$. In Fig. \ref{triple} (c) we study the dependence of $E_2^{opt}$ as a function of $T_h$, and conclude that $E_2^{opt}$ is always finite (for finite $T_h$). Hence, there appears to be an optimum value of the energy $E_2$ (that effectively sets the largest possible energy gap $E_1+E_2$ in the fridge) for which the heat current is maximized. This effect can be intuitively understood as follows: for the ideal operation of the machine, the sink reservoir must be primed to accept as much heat as possible, and thus the sink qubits must be as biased towards the ground state as much as possible, that is $T_r\ll(E_1+E_2)$. On the other hand, the hot reservoir must be primed to inject heat into the fridge, and thus the hot qubits must be as close as possible to the maximally mixed state, that is $(E_2-E_1)\ll T_h$. Hence for a given set of resources ($T_h$ and $T_r$), the heat current will be maximized when the above two conditions are both satisfied, and thus $E_2^{opt}$ is clearly always finite (for finite $T_h$).

Finally, since our quantum fridge contrasts with previous designs, we would like to compare its performance to other schemes. In particular, we compare here our model to the simplest and well-studied three-level (or qutrit) fridge, discussed in \cite{schulz,linden10}. This model also works by selectively coupling various energy transitions to three different thermal reservoirs. Analogous to our quantum fridge, a master equation and steady state of operation can also be derived for the qutrit fridge (see \cite{linden10} for details). In order to make a fair comparison, we take the resources for both the qutrit fridge and our fridge to be identical, i.e. the temperatures $T_h$ and $T_r$ of the hot and sink reservoirs are the same, and the thermal qubits from each reservoir that are connected to either fridge have the same energy gap (see Fig. \ref{sketch}). In this way, both models feature the same virtual temperature $T_v$, and thus also an identical Carnot point of operation, and identical conditions for cooling to be achieved. 

To compare the performance of the fridges away from the Carnot point, we must discuss the coupling to the baths. Here we use the collision model discussed above for both models, in which the coupling strength between a reservoir and the fridge is proportional to the rate of collisions of reservoir qubits with the fridge. For the coupling to the sink and hot reservoirs, we consider the same couplings $p_h$ and $p_r$ for both models. This is justified by the fact that the sink and hot reservoirs are connected to the same energy gaps in both models. To describe the coupling to the cold bath, the situation is a bit more subtle. While the qutrit fridge is coupled to the cold bath with a transition of energy $2E_1$, our model is connected to a transition at energy $E_1$. For the purpose of this comparison, we take $p_c$ to be the same for both our fridge and the qutrit.\footnote{The exact relationship is very much dependent upon the nature of the thermal reservoir}

The comparison between the two models is illustrated in Fig. \ref{triple}(a) and (b); here we have used the same parameters as above, in our case study. Clearly, our fridge outperforms the qutrit fridge in this example. More generally, when we conducted a numerical search over all parameters (temperatures, couplings and energies) and found regimes where our fridge outperforms the qutrit (as above), as well as regimes where the qutrit outperforms our fridge.

\section{Discussion}

We have discussed a small quantum absorption fridge. We first discussed the physics of the fridge in general terms, characterizing the cooling regime and showed that the model can reach Carnot efficiency. These conclusion were then verified using a detailed model for the fridge. Conceptually, the main interest of this model is that coupling are reversed compared to previous models. Instead of creating a virtual qubit at a cold temperature and placing it in thermal contact with the qubit to be cooled, our model refrigerates a physical qubit directly. Hence the roles of the real and virtual qubits is switched compared to standard quantum absorption fridges. 

From a more practical point of view, we found that our model performs comparably to the standard three-level quantum absorption refrigerator, in terms of cooling power, and outperforms it in some regimes. This suggests that our model is relevant for experimental implementations of quantum thermal machines.

\section{Acknowledgements} We thank Gonzalo Manzano and Sandu Popescu for discussions. We acknowledge financial support from the European project ERC-AD NLST, the Swiss National Science Foundation (grant PP00P2\_138917), SEFRI (COST action MP1006), and the EU SIQS.

\begin{appendix}
\section{Appendix}
\subsection{Bosonic heat bath}
In this section we will give the Master equation and corresponding steady state solution that corresponds to the system being in contact with bosonic thermal baths. As is well known \cite{breuer,correa2} the Master equation which follows from modelling the weak interaction between a system and a collection of Bosons with the Born and Markov approximations is given by
\begin{multline}\label{e:master bose}
\frac{d \rho}{dt} = i[\rho, H_0] + \sum_\alpha \Big(\Gamma_\alpha \big( \sigma_-^\alpha \rho \sigma_+^\alpha - \tfrac{1}{2}\{ \sigma_+^\alpha \sigma_-^\alpha , \rho \}\big) \\ + \Gamma_{-\alpha}\big( \sigma_+^\alpha \rho \sigma_-^\alpha - \tfrac{1}{2}\{ \sigma_-^\alpha \sigma_+^\alpha , \rho \}\big)\Big)
\end{multline}
Here $\alpha = c$, $r$, $h$ labels the bath, $\Gamma_{\pm\alpha}$ are the decay rates, given by $\Gamma_\alpha = \gamma_\alpha E_\alpha^3(1+N_\alpha(E_\alpha))$, where $\gamma_\alpha$ are constants of each bath (which in principle can be different, depending on how strongly the system is coupled to the bath), $E_\alpha$ is the energy coupled to the bath, $N_\alpha(E_\alpha) = 1/(e^{\beta_\alpha E_\alpha}-1)$ is the average number of excitations in the bath at temperature $\beta_\alpha$ and energy $E_\alpha$ and $\Gamma_{-\alpha} = e^{-\beta_\alpha E_\alpha} \Gamma_\alpha$.  Furthermore, $\sigma_+^\alpha$ and $\sigma_-^\alpha = {\sigma_+^{\alpha}}^\dagger$ are the `jump' operators, given by
\begin{align}
\sigma_+^c &= \ket{1}\bra{0}\otimes \hat{\mathbb{I}} \nonumber \\
\sigma_+^h &= \ket{01}\bra{10} \\
\sigma_+^r &= \ket{11}\bra{00}  \nonumber
\end{align}
where the difference between the cold bath and the room and hot baths comes from the fact that in the former case we couple to the first qubit itself, whereas in the latter cases we couple only to single transitions in the (combined) four-level system. 

Solving this equation we again find a diagonal steady state, now given by
\begin{align}\label{e:steady state bose}
	\rho_S &=  \frac{1}{\Delta} \mathrm{diag} \begin{pmatrix}
				\cg^2\rg\hg \left( \frac{1}{\cG} + \frac{1}{\hG} \right) + \cg\ce\rg \left( \frac{\he}{\cG} + \frac{\hg}{\hG} \right) \\
				\cg\ce\rg\he \left( \frac{1}{\cG} + \frac{1}{\rG} \right) + \cg^2\re \left( \frac{\he}{\cG} + \frac{\hg}{\hG} \right) \\
				\ce\rg\hg \left( \frac{\cg}{\cG} + \frac{\ce}{\hG} \right) + \cg\hg \left( \frac{\cg\re}{\cG} + \frac{\ce\rg}{\rG} \right) \\
				\cg\re\hg \left( \frac{\cg}{\cG} + \frac{\ce}{\hG} \right) + \ce\he \left( \frac{\cg\re}{\cG} + \frac{\ce\rg}{\rG} \right)
			\end{pmatrix}, 
\end{align}
where
\begin{multline}
\Delta = \cg\re\hg\left(\frac{\cg}{\cG} + \frac{\ce}{\hG}\right) + \ce\he\left(\frac{\cg\re}{\cG}+\frac{\ce\rg}{\rG}\right).
\end{multline}

Interestingly, we notice that if we make the substitution $\Gamma_\alpha \to p_\alpha r_\alpha$ then the steady state solution \eqref{e:steady state bose} to the bosonic bath Master equation \eqref{e:master bose} can be seen to exactly coincide with the steady state solution \eqref{e:steady state} to the Master equation \eqref{master} in the main text (after a small amount of algebra). Moreover, given that it is possible to tune the $\gamma_\alpha$ (the coupling strengths to the bosonic baths) it follows that one can choose the $\gamma_\alpha$ such that this mapping is satisfied for any specific choice of Hamiltonian $H_0$. 

In conclusion, we see that the two Master equations \eqref{master} and \eqref{e:master bose} describe the same physics, as far as the stationary behaviour of the fridge is concerned. 
\end{appendix}

\end{document}